\documentclass[12pt,preprint]{aastex}

\shorttitle{Outflow-Triggered Star Formation in OMC-2 FIR 3/4}
\shortauthors{Shimajiri et al.}

\begin{document}


\title{Millimeter- and Submillimeter-Wave Observations\\
of the OMC-2/3 Region. II.\\
Observational Evidence for Outflow-Triggered Star Formation\\
in the OMC-2 FIR 3/4 Region}

\author{YOSHITO SHIMAJIRI}
\affil{Department of Astronomy, School of Science, University of Tokyo, Bunkyo, Tokyo 113-0033, Japan; yoshito.shimajiri@nao.ac.jp}

\author{SATOKO TAKAHASHI$^1$}
\affil{Department of Astronomical Science, The Graduate University for Advanced Studies,
National Astronomical Observatory of Japan, Osawa 2-21-1, Mitaka, Tokyo 181-8588, Japan}

\author{SHIGEHISA TAKAKUWA}
\affil{Academia Sinica Institute of Astronomy and Astrophysics, P.O. Box 23-141, Taipei 106, Taiwan}

\author{MASAO SAITO}
\affil{ALMA Project Office, National Astronomical Observatory of Japan, Osawa 2-21-1, Mitaka, 
Tokyo 181-8588, Japan}

\and

\author{RYOHEI KAWABE}
\affil{Nobeyama Radio Observatory, Minamimaki, Minamisaku, Nagano 384-1805, Japan}


\altaffiltext{1}{Current Address: Academia Sinica Institute of Astronomy and Astrophysics,
P.O. Box 23-141, Taipei 106, Taiwan}


\begin{abstract}
We have carried out millimeter interferometric observations of the Orion Molecular Cloud-2 (OMC-2) FIR 3/4 region at an angular resolution of $\sim$ 3$\arcsec$ - 7$\arcsec$ with the Nobeyama Millimeter Array (NMA) in the H$^{13}$CO$^{+}$ ($J$=1--0), $^{12}$CO ($J$=1--0), SiO ($v$=0, $J$=2--1), and CS ($J$=2--1) lines and in the 3.3 mm continuum emission. Submillimeter single-dish observations of the same region have also been performed with Atacama Submillimeter Telescope Experiment (ASTE) in the $^{12}$CO ($J$=3--2) and CH$_3$OH ($J_K$=7$_K$--6$_K$) lines. Our NMA observations in the H$^{13}$CO$^{+}$ emission have revealed 0.07 pc-scale dense gas associated with FIR 4. The $^{12}$CO ($J$=3--2,1--0) emission shows high-velocity blue and red shifted components at the both north-east and south-west of FIR 3, suggesting a molecular outflow nearly along the plane of the sky driven by FIR 3. The SiO ($v$=0, $J$=2--1) and the submillimeter CH$_{3}$OH ($J_K$=7$_K$--6$_K$) emission, known as shock tracers, are detected around the interface between the outflow and the dense gas. Furthermore, the $^{12}$CO ($J$=1--0) emission shows an L-shaped structure in the P-V diagram. These results imply presence of the shock due to the interaction between the molecular outflow driven by FIR 3 and the dense gas associated with FIR 4. Moreover, our high angular-resolution ($\sim$ 3$\arcsec$) observations of FIR 4 in the 3.3 mm continuum emission with the NMA have first found that FIR 4 consists of eleven dusty cores with a size of $\sim$ 1500 - 4000 AU and a mass of $\sim$ 0.2 - 1.4 M$_{\odot}$. The separation among these cores ($\sim$ 5 $\times 10^3$ AU) is on the same order of the Jeans length ($\sim$ 13 $\times 10^3$ AU), suggesting that the fragmentation into these cores has been caused by the gravitational instability. The time scale of the fragmentation ($\sim$ 3.8 $\times 10^4$ yr), estimated from the separation divided by the sound speed, is similar to the time scale of the interaction between the molecular outflow and the dense gas ($\sim$ 1.4 $\times 10^4$ yr).  We suggest that the interaction between the molecular outflow from FIR 3 and the dense gas at FIR 4 triggered the fragmentation into these dusty cores, and hence the next generation of the cluster formation in FIR 4.

\end{abstract}





\section{Introduction}
The process of so-called ``isolated'' star formation, where a single star or binary stars form in one molecular cloud core, has been studied intensively \citep{Sai99,Sai01,tak04,Jes07,tak07a}. In this mode, spontaneous gravitational collapse of cloud cores initiates the formation of the protostar at the center, and there is observational evidence of such gravitational collapse \citep{Myers95,Sai96,Mar97,Oha97,Momo98,tak07b}. Another mode of star formation is cluster formation, where stellar clusters form simultaneously or sequentially in one cloud core. A vast majority ($\geq$ 90 \%) of stars form as members of clusters in our galaxy \citep{Lada03}, and the understanding of the cluster formation is crucial for more comprehensive understanding of star formation. Previous observational studies of cluster formation have proposed that external effects on cloud cores, such as stellar wind, HII regions \citep{Wil94,Wil95,Hester05}, supernova \citep{koba05,Yasu06}, and molecular outflows \citep{San01,Nakano03}, are often required to trigger cluster formation. Theoretical studies also suggest that such external effects are often required to trigger cluster formation \citep{Norman80}. For example, \citet{Yasu06} have found 52 cluster members in ``Cloud 2'', which locates in the vicinity of the SNR H I shell, GSH 138-01-94. The age of the cluster in Cloud 2 ($\sim$ 1.0 $\times$ 10$^6$ yr) is younger than that of GSH 138-01-94 ($\sim$ 4.3 $\times$ 10$^6$ yr), and they suggest that the star formation in Cloud 2 was triggered by the interaction with the SNR H I shell. In the 850 $\mu$m dust continuum map of the NGC 1333 region, there are several cavity and/or shell structures observed, which are most likely created by molecular outflows \citep{San01}. One dusty core, SK-1, is in one of these shells and is gravitationally bound, and  \citet{San01} suggest that this core is a specific example of the outflow-triggered star formation. However, these studies are presumptions mainly based on the overall morphology, and detailed studies of cluster formation processes, including spatially-resolved observations of both triggers and cloud cores and the investigation of the physical processes, have been limited. This is mainly due to the relatively far distance to cluster-forming regions and lack of interferometric observations of cluster-forming regions in millimeter and submillimeter molecular lines as well as continuum emissions. 

 In order to investigate the detailed mechanism of cluster formation, we have initiated millimeter and submillimeter survey observations of the OMC-2/3 region with the Nobeyama Millimeter Array (NMA) and Atacama Submillimeter Telescope Experiment (ASTE). The OMC-2/3 region is one of the nearest cluster-forming regions at a distance of 450 pc \citep{Gen89,Johnson90,Lada03,nie03}, and locates at the north of the Orion A Giant Molecular Cloud \citep{Gat74,Sak94,Chini97,lis98}. Mapping observations of the OMC-2/3 region in the 1.3 mm continuum emission have identified 26 dusty cores \citep{Chini97,nie03}, and 33 dusty cores in the 850 $\mu $m emission \citep{lis98}. $^{12}$CO ($J$=1--0) survey observations of molecular outflows in OMC-2/3 have found nine outflows at an average separation of $\sim$ 0.2 pc \citep{Aso00,wil03}, and our recent submillimeter $^{12}$CO ($J$=3--2) survey observations of OMC-2/3 have revealed presence of 15 outflows \citep{Takaha07}. Extensive mapping observations of the OMC-2/3 region in the CS ($J$=1--0) \citep{Tate93} and H$^{13}$CO$^{+}$ ($J$=1--0) lines \citep{Aso00, ike07} have unveiled physical properties of cloud cores in the cluster-forming region, which are quantitatively different from those in isolated star-forming regions. Interferometric H$^{13}$CO$^{+}$($J$=1--0) observations of MMS 7 in the OMC-3 region, one of the dusty cores, have revealed the disk-like envelope around MMS 7, which is being dispersed by the associated molecular outflow \citep{Takaha06}. 

 In this paper, we focus on observations of the OMC-2 FIR 3/4 region 
\citep[including FIR 3, 4, and 5;][]{Chini97}
FIR 4 is the strongest 1.3 mm dust-continuum source in OMC-2 \citep{Joho99,Chini97}. In this region, there are three 3.6 cm free-free emission sources of VLA 11 (consistent with FIR 3), VLA 12 and VLA 13 \citep{rei99}, and nine MIR sources \citep{nie03}, suggesting that FIR 4 is an active cluster-forming region. A molecular outflow driven by FIR 3 has been found in the $^{12}$CO ($J$=1--0, 3--2) lines \citep{wil03,Takaha07}, which is one of the candidate triggers of the cluster formation in FIR 4. Our millimeter interferometric and submillimeter single-dish observations of the FIR 4 region should shed light on the detailed mechanism of the cluster formation, which will be shown in the present paper. 
 
\section{Observations and Data Reduction}
\subsection{NMA observations}

We carried out millimeter interferometric observations of the FIR 4 region in the H$^{13}$CO$^{+}$ ($J$=1--0; 86.754 GHz), $^{12}$CO ($J$=1--0; 115.271 GHz), SiO ($v$=0, $J$=2--1; 86.847 GHz), and CS ($J$=2-1; 97.980 GHz) lines with the Nobeyama Millimeter Array (NMA), which consists of six 10 m antennas, during a period from 2005 May to 2007 January. The data in the H$^{13}$CO$^{+}$ ($J$=1--0), $^{12}$CO ($J$=1--0), and SiO ($v$=0, $J$=2--1) lines were obtained with the FX correlator, which was configured with 1024 channels per baseline and a bandwidth of 32 MHz. For the $^{12}$CO (1--0) and SiO (2--1) data we made 5-channel binning to increase the signal-to-noise ratio ($\equiv$ S/N) of the high-velocity line-wing emission. Thus, the velocity resolution in the H$^{13}$CO$^{+}$, $^{12}$CO, and SiO observations is 0.108, 0.406, and 0.539 km s$^{-1}$, respectively. The CS ($J$=2--1) data were obtained with the digital spectral correlator, Ultra Wide Band Correlator (UWBC) \citep{oku00}, which has 128 channels and a 1024 MHz bandwidth per baseline (1 channel = 24.47 km s$^{-1}$). The hanning window function was applied to reduce side lobes in the sampling. Thus, the effective velocity resolution was widened to 48.94 km s$^{-1}$ (corresponding to 2 channels). The CS ($J$=2--1) emission is detected over the three UWBC channels with the hanning window function, and we integrated over these three channels (corresponding to 97.88 km s$^{-1}$) to make the total integrated intensity map. We also obtained continuum data at both the lower (87.090 $\pm $ 0.512 GHz) and upper (98.418 $\pm $ 0.512 GHz) sidebands with UWBC, simultaneously with the H$^{13}$CO$^{+}$ ($J$=1--0) data. To obtain a higher S/N in the continuum map, the data of both sidebands were co-added (effective observing frequency = 92 GHz = 3.3 mm). Using the AIPS package developed at NRAO, we adopted both the uniform and natural UV weighting for the continuum imaging. For the imaging of the molecular emissions we adopted the natural weighting. Table \ref{NMAobs} and \ref{continuumobs} summarize the parameters for the line and continuum observations, respectively. Since the minimum projected baseline length of the H$^{13}$CO$^{+}$ ($J$=1--0), $^{12}$CO ($J$=1--0), SiO ($v$=0, $J$=2--1), and CS ($J$=2--1) observations was 2.9, 3.0, 2.9, and 3.2 k$\lambda$, our observations were estimated to be insensitive to structures more extended than 57$\arcsec$ (0.13 pc), 55$\arcsec$ (0.12 pc), 57$\arcsec$ (0.13 pc), and 51$\arcsec$ (0.11 pc) at the 10 \% level, respectively \citep{Wilner94}. The overall uncertainty in the flux calibration was estimated to be $\sim$ 15 \%. After the calibrations, only the data taken under good weather conditions were adopted in the imaging. 

\subsection{ASTE observations}

The submillimeter CH$_3$OH ($J_K$=7$_K$--6$_K$; K=-1, 0, 2; 338.408681, 338.344629, 338.72165 GHz) data in the FIR 4 region have been taken with the ASTE 10 m telescope (Atacama Submillimeter Telescope Experiment; \citet{eza05,Koh04}) located at the Pampa la Bola (altitude = 4800 m), Chile.  The observed frequency range covers the CH$_{3}$OH K-ladder (7$_K$--6$_K$, $K$=-2, -1, 0, 2, and 6) from 338.32 GHz to 338.74 GHz, where three of those (7$_{-1}$--6$_{-1}$, 7$_0$--6$_0$, 7$_2$--6$_2$) are detected above 3 $\sigma$ level (= 0.51 K). Observations were remotely made from the ASTE operation room at Mitaka, Tokyo on 2006 October 7, 10, and 11, using the network observation system N-COSMOS3 developed by NAOJ \citep{kam05}. The half-power beam width of the ASTE telescope is 22$\arcsec$ at the observing frequency. The typical system noise temperature with the 345 GHz SIS heterodyne receiver in the DSB mode was 270 - 440 K, and the typical atmospheric opacity at 220 GHz was $\sim$ 0.05 during our observations. The temperature scale was determined by the chopper-wheel method, which provides us with the antenna temperature corrected for the atmosphere attenuation. As a backend, we used four sets of a 1024-channel auto-correlator, which provide us a frequency resolution of 31.12 kHz, corresponding to 0.1 km s$^{-1}$ at the CH$_3$OH ($J_K$=7$_K$--6$_K$) frequency. The On-The-Fly (OTF) mapping technique was employed to map the $5\arcmin \times 5\arcmin$ region centered on FIR 4, which covers the entire outflow driven by FIR 3 seen in the $^{12}$CO (3--2) emission \citep{Takaha07}. The telescope pointing was checked every two hours by 5-points scans of the point-like $^{12}$CO ($J$=3--2) emission from O-Cet ($\alpha _{J2000}$= 02$^{h}$ 19$^{m}$ 20$\fs$80 $\delta _{J2000}$= -02$^\circ$ 08$\arcmin$ 40$\farcs$7). The pointing errors were measured to be from 0$\farcs$2 to 1$\farcs$2 during the observing run. We also observed NGC 1333 IRAS 2A and compared the observed CH$_{3}$OH spectra to those obtained with the Caltech Submillimeter Observatory (CSO) telescope, which has the same dish size as that of the ASTE telescope \citep{Jes04}, and found that the main beam efficiency of the ASTE telescope was $\sim$ 0.4. After subtracting linear baselines, the OTF data were convolved by a Gaussian-tapered Bessel function \citep{man00} with a FWHM of 14$\arcsec$ and were resampled onto a 7$\arcsec$ grid. Since the telescope beam is a gaussian with a FWHM of 22$\arcsec$, the effective FWHM resolution in the restored images is 26$\arcsec$. The ``scanning effect" was minimized by combining scans along the R.A. and decl. direction, using the PLAIT algorithm developed by \cite{Eme88}. 
The typical rms noise level in the final image is 0.17 K in $T_A^*$ at a velocity resolution of 0.5 km s$^{-1}$.

\section{RESULTS}
\subsection{$^{12}$CO ($J$=3--2, 1--0) emission}
\begin{figure}
  \includegraphics[width=11cm,clip]{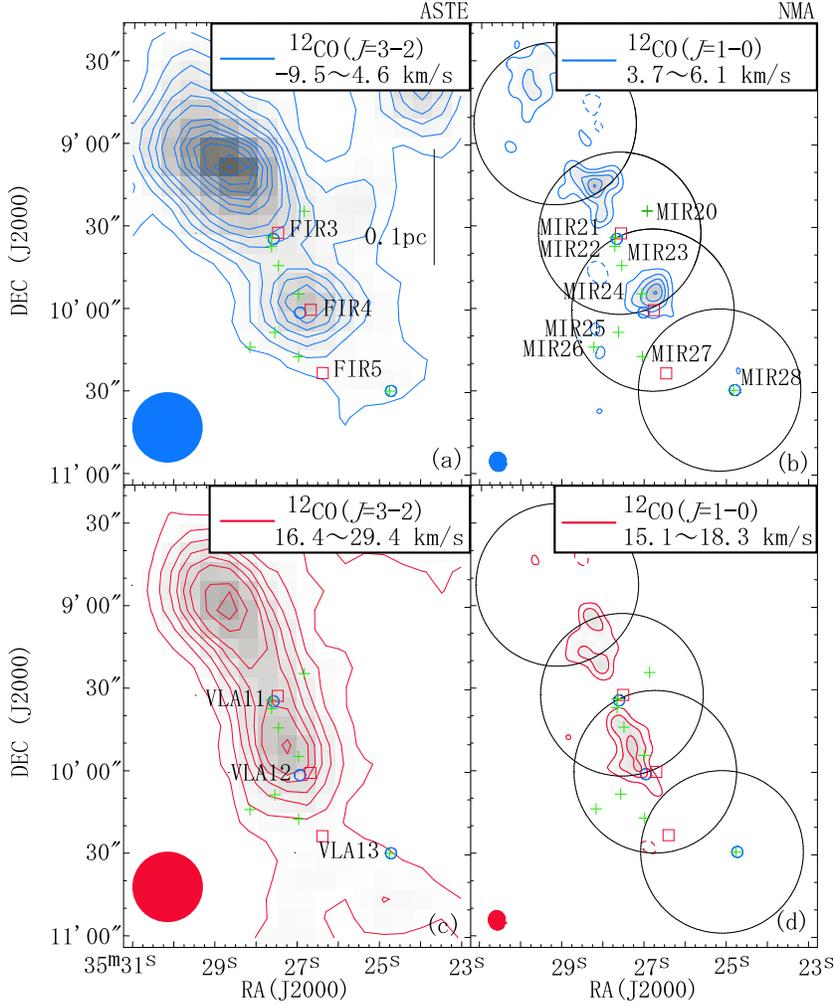}
  \caption{Distribution of the high-velocity blueshifted (blue contour) and redshifted (red contour)
$^{12}$CO ($J$=3--2) (left) and $^{12}$CO ($J$=1--0) (right) emission in the OMC-2/FIR 4 region.
Panel (a) and (c) show the distribution of the blue (-9.5 km s$^{-1}$ - 4.6 km s$^{-1}$)
and red (16.4 km s$^{-1}$ - 29.4 km s$^{-1}$) lobe in the $^{12}$CO ($J$=3--2)
line observed with ASTE \citep{Tahaka07}, while
panel (b) and (d) show the distribution of the blue (3.7 km s$^{-1}$ to 6.1 km s$^{-1}$)
and red (15.1 km s$^{-1}$ to 18.3 km s$^{-1}$) lobe in the $^{12}$CO ($J$=1--0)
line observed with the NMA, respectively.
Open black circles show the field of view of the NMA observations, green crosses
the position of MIR sources \citep{nie03}, blue circles the position of 3.6 cm free-free sources
\citep{rei99}, and red squares the peak position of the 1.3 mm dust-continuum emission \citep{Chini97}. Filled
ellipses at the bottom left corner of each panel indicate the beam size.  Contour levels of these maps start at
$\pm$ 5 $\sigma$ levels with an interval of 5 $\sigma$. The rms noise levels (1 $\sigma$) of the panel (a), (b),
(c) and  (d) are 1.2 K, 0.45 Jy, 1.2 K, and 0.70 Jy beam$^{-1}$, respectively.}
\label{Line_map_CO}
\end{figure}

\begin{figure}
  \includegraphics[width=15cm,clip]{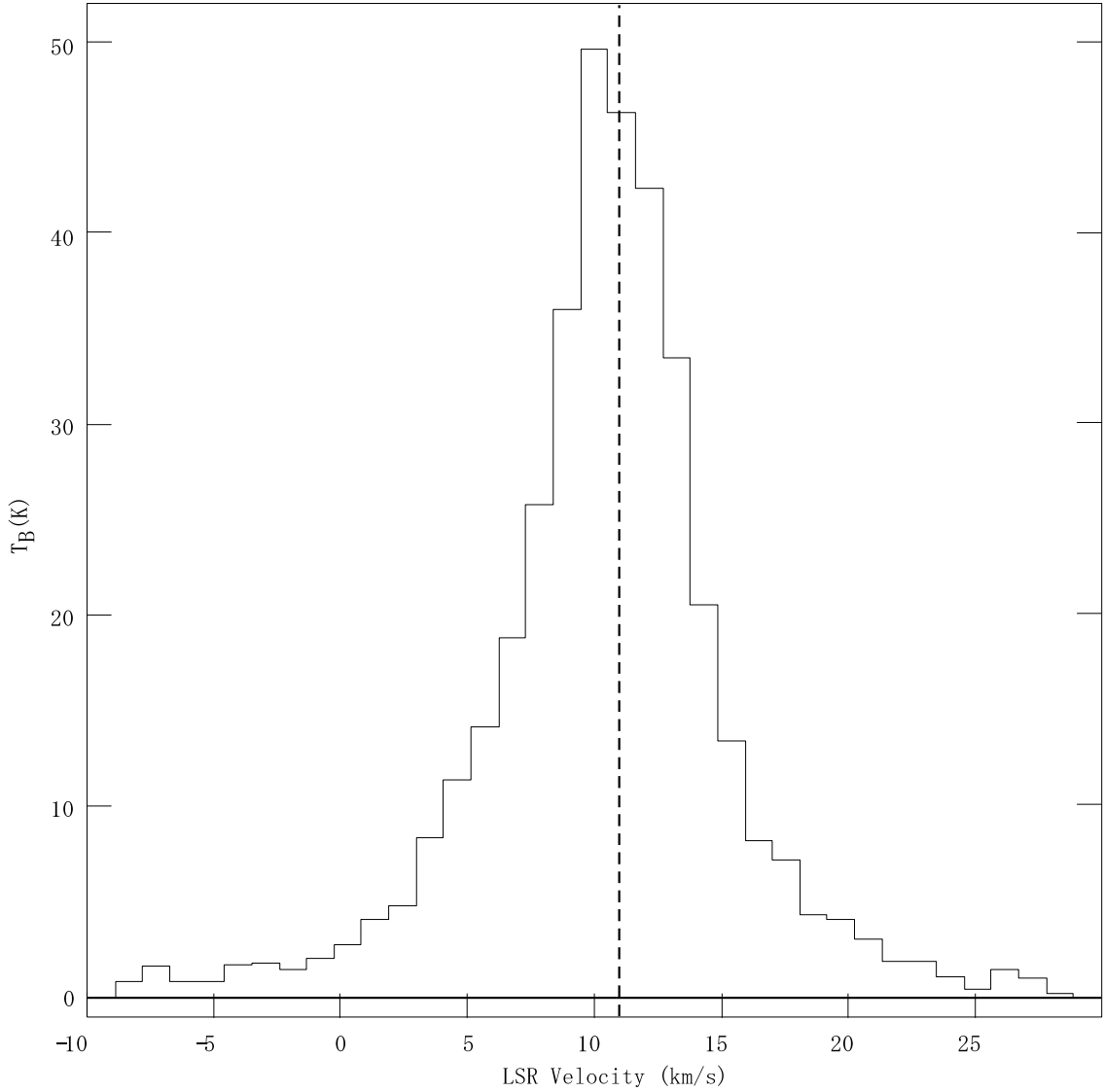}
  \caption{$^{12}$CO ($J$=3--2; 345.705 GHz) spectrum at FIR 4 taken with ASTE \citep{Takaha07}
The rms noise level (1$\sigma $) is 0.5 K in T$_B$ at a velocity resolution of 1.1 km s$^{-1}$.
A dashed line shows the systemic velocity ($\sim$ 11.3 km s$^{-1}$), obtained from the Gaussian fit to the spectrum.}
  \label{12cospec}
\end{figure}

 Figure \ref{Line_map_CO} shows the distribution of the high-velocity blueshifted and redshifted $^{12}$CO ($J$=3--2) and $^{12}$CO ($J$=1--0) emission in the FIR 4 region, observed with ASTE and NMA, respectively. The $^{12}$CO ($J$=3--2) data were taken by our group as a part of our survey project in the OMC-2/3 region \citep{Takaha07}. The $^{12}$CO ($J$=3--2) line profile toward FIR 4 is shown in Figure \ref{12cospec}. From the peak temperature of the $^{12}$CO ($J$=3--2) emission, we determine that the gas kinetic temperature in the FIR 4 region is $\sim$ 52.5 K. The detected velocity range of the $^{12}$CO ($J$=3--2) and  $^{12}$CO ($J$=1--0) emission is -9.5 $\sim$ 29.4 km s$^{-1}$ and 3.7 $\sim$ 18.3 km s$^{-1}$, respectively, while that of the H$^{13}$CO$^{+}$ ($J$=1--0) emission, which traces the dense gas in FIR 4, is 10.4 $\sim$ 12.5 km s$^{-1}$ \citep{ike07}. There are both blueshifted and redshifted high-velocity components at the north-east (NE) and south-west (SW) of FIR 3. These results imply that the high-velocity components in the $^{12}$CO ($J$=3--2, 1--0) emission trace a bipolar molecular outflow nearly along the plane of the sky driven by FIR 3. Furthermore, the high-velocity components of the $^{12}$CO emission at the north-east of FIR 3 are associated with H$_{2}$ knots found by \citet{Yu97}, which also supports our outflow interpretation. In addition, the velocity range of the $^{12}$CO (3--2) emission is larger than that of the $^{12}$CO (1--0) emission. One of the plausible interpretations of the different velocity range is that the submillimeter line traces warmer gas in the vicinity of the jet, where the higher velocity is expected, due to the different excitation condition between the submillimeter and millimeter lines \citep{Raga93}.

The size of the SW blue lobe measured from the $^{12}$CO ($J$=3--2) image is $\sim$ 0.09 pc, which is twice smaller than that of the NE lobe ($\sim$ 0.19 pc). At the south-west of FIR 3 both the $^{12}$CO ($J$=3--2) and $^{12}$CO ($J$=1--0) emission exhibit blueshifted and redshifted peaks near FIR 4, and that these ASTE and NMA peaks appear to be consistent with each other. On the other hand, at the north-east of FIR 3 the prominent blueshifted and redshifted peaks observed with ASTE are not seen in the NMA maps, suggesting that these components are resolved out by the interferometer. These results imply that the SW component of the outflow driven by FIR 3 toward FIR 4 is more compact than the NE component. From our NMA observations, the peak position of the SW red lobe is measured to be $\alpha _{J2000}$=05$^{h}$ 35$^{m}$ 27$\fs$2, $\delta _{J2000}$=-05$^\circ$ 09$\arcmin$ 59$\farcs$5, and at the tip of the red lobe there is 3.6 cm free-free emission (VLA 12, open blue circle). There are two possibilities on the origin of this free-free emission. One is that there is another protostar located at VLA 12, and the other is that this free-free emission is originated from the ionized jet driven by FIR 3. Since there is no 3.3-mm dust continuum counterpart associated with VLA 12 (see $\S$ 3.4), the origin of the 3.6 cm free-free emission is probably the ionized jet driven by FIR 3. 

\begin{figure}
  \includegraphics[width=15cm,clip]{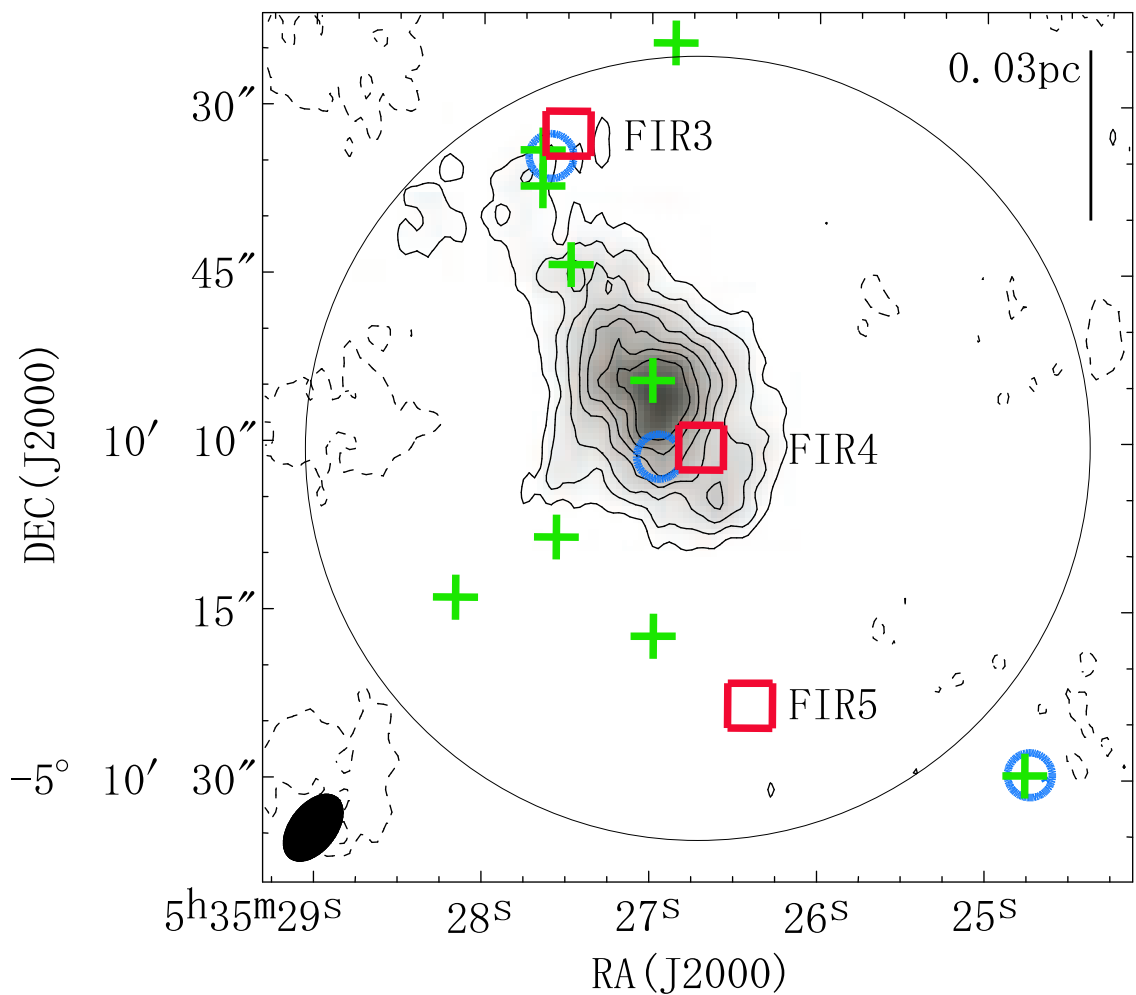}
  \caption{Total integrated intensity map of the CS ($J$=2--1) emission in the OMC-2/FIR 4 region, obtained with the NMA.
Symbols in the Figure are the same as in Figure \ref{Line_map_CO}.
Contour levels of this map start at $\pm$ 3 $\sigma$ levels with an interval of 1 $\sigma$.
The rms noise level (1 $\sigma$) of this map is 5.1$\times$10$^{-3}$ Jy beam$^{-1}$. }
  \label{CS}
\end{figure}

Figure \ref{CS} shows a total integrated intensity map in the CS ($J$=2--1) emission. The peak of this emission is ($\alpha_{J2000}, \delta_{J2000} $) = (05$^{h}$ 35$^{m}$ 27$\fs$0, -05$^\circ$ 09$\arcmin$ 56$\farcs$0), $\sim$ 4$\arcsec$ north-east of FIR 4. This emission extends from FIR 3 to FIR 4, similarly to the high-velocity $^{12}$CO ($J$=1--0, 3--2) emission shown in Figure \ref{Line_map_CO}. Thus, one of the possible interpretations on the origin of the CS emission is moderately dense gas in the molecular outflow. Higher velocity-resolution observations in the CS emission are required to verify the origin of the CS emission in the FIR4 region.

\subsection{Dense gas associated with FIR 4}
\begin{figure}
  \includegraphics[width=15cm,clip]{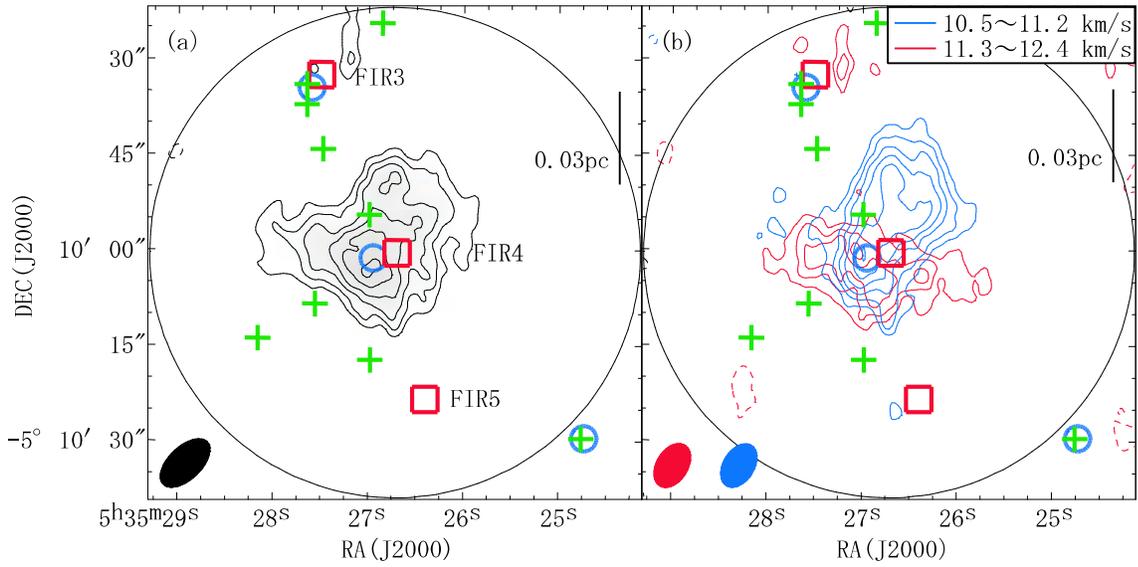}
  \caption{Total integrated intensity map of the H$^{13}$CO$^{+}$ ($J$=1--0) emission (left), and maps of the
blueshifted (10.5 - 11.2 km s$^{-1}$) the redshifted (11.3 - 12.4 km s$^{-1}$)
H$^{13}$CO$^{+}$ components (right) in the FIR 4 region, obtained with NMA. 
Symbols in the Figure are the same as in Figure \ref{Line_map_CO}.
Contour levels of these maps start at $\pm$ 3 $\sigma $ levels with an interval of 1 $\sigma$.
The rms noise levels (1 $\sigma$) in panel (a), the blue and red contour in panel (b) 
are 0.81, 0.49 and 0.49 Jy beam$^{-1}$, respectively. }
  \label{H13CO+}
\end{figure}

Figure \ref{H13CO+}a shows a total integrated intensity map of the H$^{13}$CO$^{+}$ ($J$=1--0) emission in the FIR 4 region, observed with the NMA. There exists a $\sim$ 0.07 pc-scale ($\sim$ 30$\arcsec$) dense gas structure associated with FIR 4. This dense gas structure has also been identified as AC17 with the single-dish 45m telescope by \citet{Aso00}. Hereafter we call this structure ``FIR 4 clump". The peak position of FIR 4 clump ($\alpha_{J2000}$=05$^{h}$35$^{m}$27$\fs$0, $\delta_{J2000}$=-05$^\circ$ 10$\arcmin$ 03$\farcs$5), estimated by the 2-dimensional Gaussian fitting to the image, is consistent with the position of AC17, and the systemic velocity ($\equiv $V$_{sys}$$\sim$ 11.3 km s$^{-1}$) and the line width ($\sim$ 1.1 km s$^{-1}$) estimated by the Gaussian fitting to the H$^{13}$CO$^{+}$ spectrum at FIR 4 are also consistent with the single-dish results \citep{Aso00}. 
Figure \ref{H13CO+}b shows the distribution of the blue- (10.5 km s$^{-1}$ - 11.2 km s$^{-1}$) and redshifted (11.3 km s$^{-1}$ - 12.4 km s$^{-1}$) H$^{13}$CO$^{+}$ ($J$=1--0) emission in FIR 4. The blueshifted component is distributed from south-east to north-west while the redshifted component is distributed from east to west.

We derived the mass of FIR 4 clump ($\equiv$ M$_{gas}$) on the assumption of the Local Thermodynamic Equilibrium (LTE) condition and the optically-thin H$^{13}$CO$^{+}$ ($J$=1--0) emission, using the following equation \citep{Takaha06},

\begin{equation}
M_{gas}=\frac{3kc^{2}{m}D^{2}}{16h\nu^{3}\pi^{3}B\mu }\frac{T_{ex}}{X[H^{13}CO^{+}]}[ exp\left( \frac{hB(J+1)(J+2)}{kT_{ex}} \right )] \int  F_{\nu } d\nu, \label{H13CO+LTE_MASS}
\end{equation}

 where $k$, $h$, $c$, $T_{ex}$, $D$, X[H$^{13}$CO$^{+}$], $\mu$, $m$, $\nu$, $J$, B and $F_{\nu }$ are the Boltzmann constant, the Planck constant, the speed of light, the excitation temperature, distance, fractional abundance of H$^{13}$CO$^{+}$, dipole moment, the mean mass of the molecule (2.33 a.m.u), frequency, rotational quantum number, rotational constant, and the flux density, respectively. We assume that the excitation temperature of the H$^{13}$CO$^{+}$ ($J$=1--0) line is equal to 52.5 K, which is derived from the peak brightness temperature of the $^{12}$CO ($J$=3--2) emission (see $\S$ 3.1). We adopt $\mu$ = 4.07 Debye as the H$^{13}$CO$^{+}$ dipole moment \citep{Hae79}. The H$^{13}$CO$^{+}$ ($J$=1--0) abundance of 4.5 $\times$ 10$^{-11}$ is adopted from \citet{Aso00}, and the derived mass of FIR 4 clump is $\sim$ 10.5 $M_{\odot}$ (If we adopt the abundance of 1.4 $\times$ 10$^{-10}$ estimated by \citet{Takaha06}, the mass of FIR 4 clump is $\sim$ 3.6 $M_{\odot}$). We also estimated the momentum ($\equiv P_{gas}$) and the internal gas energy ($\equiv  E_{gas}$) of FIR 4 clump as follows \citep{tak03a} ;  

\begin{equation}
P_{gas}=M_{gas}C_{eff}, \label{P_gas}
\end{equation}

\begin{equation}
E_{gas}=\frac{1}{2}M_{gas}C_{eff}^{2}, \label{E_gas}
\end{equation}

\begin{equation}
C_{eff}^2=\frac{\Delta V_{FWHM}^2}{8\ln2}+kT_K(\frac{1}{m}-\frac{1}{m_{obs}}), \label{C_eff}
\end{equation}

where $\Delta V_{FWHM}$, $C_{eff}$, T$_{K}$, and m$_{obs}$ is the FWHM velocity width of the H$^{13}$CO$^{+}$ line, the effective sound speed, the gas kinetic temperature, and the mass mean molecular of the observed molecule. Table \ref{Physical_qua} summarizes our estimates of these parameters. Due to the effect of the missing flux in the interferometric observations ($\sim$ 83 \% of the total flux is missed), these values should be considered as a lower limit, and in fact, the estimated mass from the single-dish data is $\sim$ 35 $M_{\odot}$ \citep{Aso00}.

\subsection{Shock tracer of CH$_3$OH ($J_K$=7$_K$--6$_K$) and SiO ($v$=0, $J$=2--1)}

\begin{figure}
  \includegraphics[width=15cm,clip]{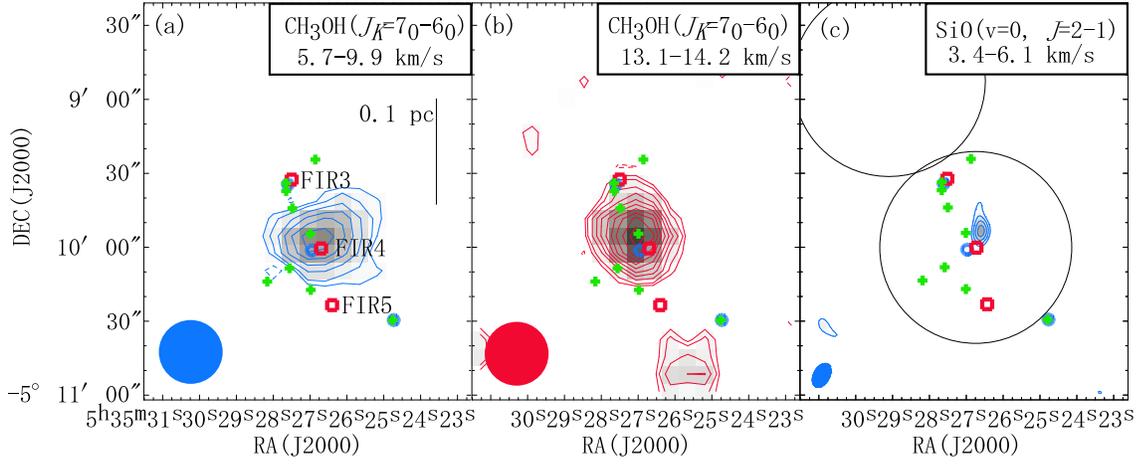}
  \caption{ASTE CH$_3$OH ($J_K$=7$_0$--6$_0$) and NMA SiO ($v$=0, $J$=2--1) maps in the FIR 4 region.
Panel (a) and (b) show the distribution of the blue (5.7 km s$^{-1}$ - 9.9 km s$^{-1}$)
and red (13.1 km s$^{-1}$ - 14.2 km s$^{-1}$) lobe in the CH$_3$OH ($J_K$=7$_0$--6$_0$) line,
respectively.
Panel (c) shows the distribution of the blue lobe in the SiO ($v$=0, $J$=2--1) line at a velocity
range from 3.4 km s$^{-1}$ to 6.1 km s$^{-1}$.
Symbols in the Figure are the same as in Figure \ref{Line_map_CO}. Contour levels in panel (a) and (b) start at
$\pm$ 5 $\sigma$ levels with an interval of 5 $\sigma$. Contour levels in panel (c) start
at $\pm$ 3 $\sigma $ levels with an interval of 1 $\sigma $.
The rms noise levels (1 $\sigma$) in panel (a), (b) and (c) are 0.024, 0.015 K, and 0.11 Jy beam$^{-1}$, respectively.}
  \label{Line_map_shock}
\end{figure}

\begin{figure}
  \includegraphics[width=15cm,clip]{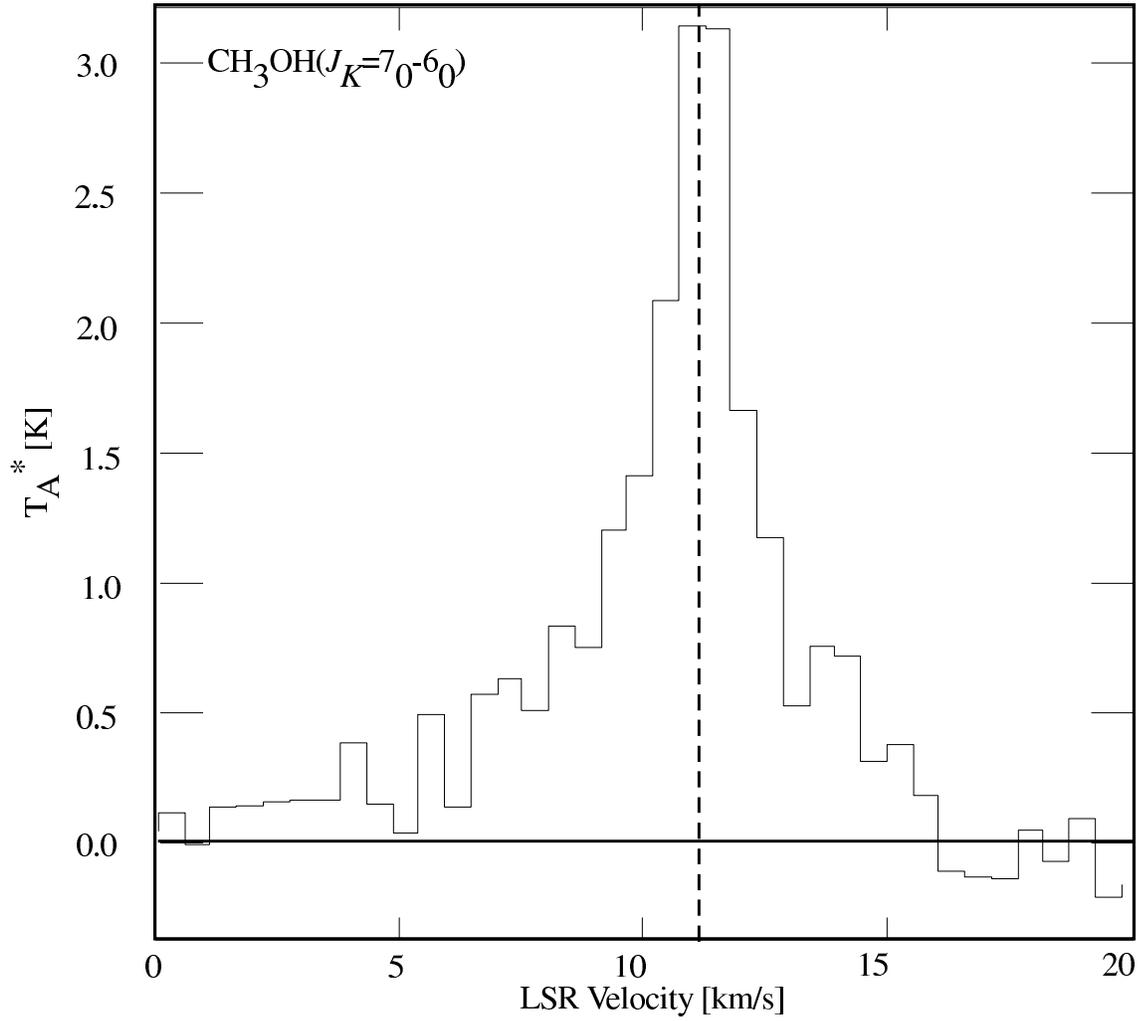}
  \caption{CH$_3$OH ($J_K$=7$_0$--6$_0$; 338.408 GHz) spectrum at FIR 4 taken with ASTE.
The rms noise level (1$\sigma$) is 0.17 K in T$_A^{\ast }$ at a velocity resolution of 0.5 km s$^{-1}$.
A dashed line shows the systemic velocity ($\sim$ 11.3 km s$^{-1}$) in the FIR 4 region estimated from
the NMA H$^{13}$CO$^{+}$ spectrum.}
  \label{spectral}
\end{figure}

We have detected the submillimeter CH$_3$OH ($J_K$=7$_K$--6$_K$; K=-1, 0, 2) lines toward FIR 4 clump with ASTE. Figure \ref{Line_map_shock} (a) and (b) show the distribution of the blue- and redshifted CH$_3$OH ($J_K$=7$_0$--6$_0$) emission, and Figure \ref{spectral} shows the spectrum of the CH$_3$OH ($J_K$=7$_0$--6$_0$) line toward FIR 4. The line width of the submillimeter line is up to 10 km s$^{-1}$ at the zeroth level, and there is slight indication of the asymmetric line profile, with a more prominent blue wing (5.7 km s$^{-1}$ - 9.9 km s$^{-1}$) than a red wing (13.1 km s$^{-1}$ - 14.2 km s$^{-1}$). The peak position of the blue and red lobe is ($\alpha_{J2000}$, $\delta_{J2000}$)=(5$^{h}$ 35$^{m}$ 26$\fs$6, -5$^\circ$ 09$\arcmin$ 55$\farcs$4) and (5$^{h}$ 35$^{m}$ 27$\fs$1, -5$^\circ$ 09$\arcmin$ 55$\farcs$4), respectively. The distribution of the blueshifted CH$_3$OH emission extends along the east-west direction, while the distribution of the redshifted emission extends toward the direction of the outflow driven by FIR 3, i.e. along SW-NE. 

We have also detected  the SiO ($v$=0, $J$=2--1) emission toward FIR 4 clump with the NMA, as shown in Figure \ref{Line_map_shock} (c). The peak position of the SiO emission is ($\alpha_{J2000}$, $\delta_{J2000}$) = (05$^{h}$ 35$^{m}$ 26$\fs$6, -05$^\circ$ 09$\arcmin$ 53$\farcs$0). The SiO emission is blueshifted, point-like, and locates at $\sim$7$\arcsec$ north of FIR 4. The peak position of the SiO emission is approximately consistent with the peak position of the blueshifted CH$_3$OH emission.

\subsection{3.3-mm dust continuum emission}
\begin{figure}
  \includegraphics[width=15cm,clip]{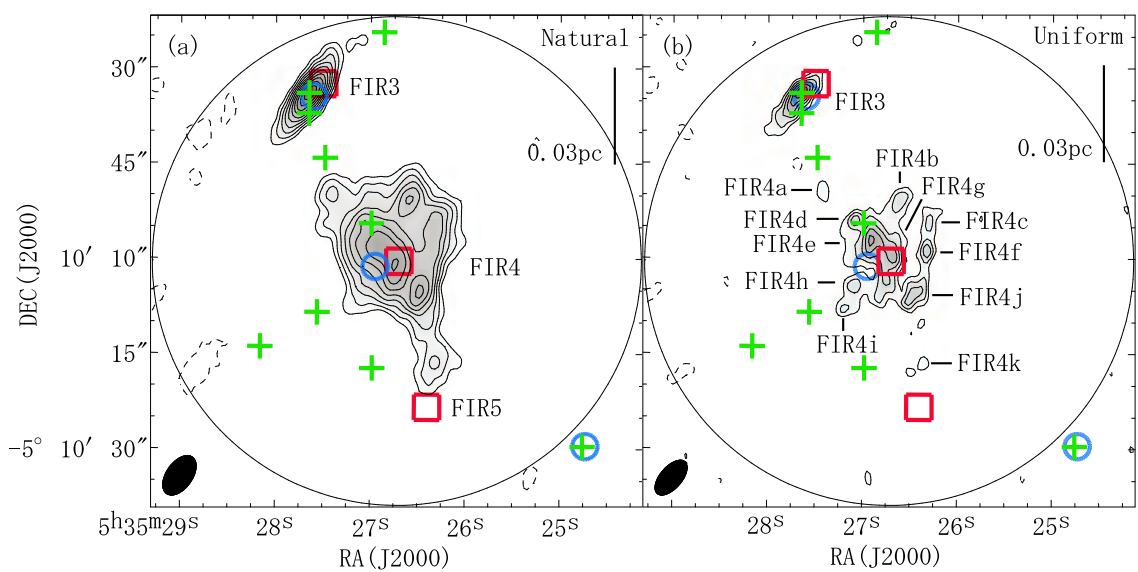}
  \caption{3.3 mm dust-continuum maps in the FIR 4 region observed with the NMA. The left panel shows
the naturally-weighted map, and the right panel the uniformly-weighted map.
Symbols in the figure are the same as in Figure \ref{Line_map_CO}. Contour levels
start at $\pm$ 3 $\sigma $ noise levels with an interval of 1 $\sigma $. The rms noise level
(1 $\sigma $) in panel (a) and (b) are 1.2$\times$10$^{-3}$ and 1.4$\times$10$^{-3}$ Jy beam$^{-1}$, respectively.
}
  \label{continuum}
\end{figure}

Figure \ref{continuum}a and \ref{continuum}b show naturally- and uniformly-weighted 3.3-mm continuum images in the FIR 4 region observed with the NMA. In these images there appears two main components, one is associated with FIR 3 and the other with FIR 4. The overall distribution is consistent with that of the single-dish 1.3 mm continuum observations \citep{Chini97}. The overall ridge distribution of the 3.3-mm emission associated with FIR 4 also resembles that of the H$^{13}$CO$^{+}$ ($J$=1--0) emission, although there appears north-eastern extension in the 3.3-mm emission. The north-eastern extension in the 3.3-mm emission may trace the ``wall" of the molecular outflow, as seen in other observations \citep{Gueth03,Mori06}. However, the high-velocity CO lobes are unlikely to be the cloud-core components traced by the H$^{13}$CO$^{+}$ emission, as discussed in section 3.1.  From these considerations, we suggest that most of the 3.3-mm emission traces the dense-gas component in FIR 4, with a contamination from the outflow component to a certain extent.

FIR 3 was detected by our 3.3 mm dust-continuum observations as a single source, although two MIR sources MIR 21 and 22, are present in FIR 3 \citep{nie03}. \citet{nie03} have suggested that MIR 21 and 22 are binary stars at a separation of $\sim$ 3$\arcsec$ (= 1350 AU). However, we consider that MIR 21 and MIR 22 are reflection nebulae evacuated by the outflow from FIR 3, because the 3.3 mm peak at FIR 3 locates between MIR 21 and MIR 22 and the alignment of MIR 21 and 22 is consistent with the direction of the molecular outflow.

The dusty component associated with FIR 4 is most likely the same identity traced by the H$^{13}$CO$^{+}$ ($J$=1--0) emission, that is, FIR4 clump. Moreover, the dusty component in FIR 4 has subpeaks. These peaks are probably substructures ($\equiv $ cores) in FIR 4 clump. We identified those cores with the following criteria; (1) the peak intensity should be higher than 4 $\sigma $ noise level, and (2) the ``valley'' among different cores should be deeper than 1 $\sigma$. We identified five cores in the naturally-weighted image, and six more cores in the uniformly-weighted image. Then the total number of the identified cores is eleven. Hereafter, we call these smaller-scale cores FIR 4a,b,\dots,k, as labeled in Figure \ref{continuum}b. 
As we mentioned in $\S$ 3.1., there is no 3.3-mm continuum source toward VLA 12. 

The deconvolved size of these dusty cores was estimated by the 2-dimentional Gaussian fitting to the image. The mass of the cores ($\equiv$ $M_{dust}$) was derived from the total 3.3 mm flux, $F_{\nu}$, on the assumption that all the 3.3-mm continuum emission arises from dust and that the emission is optically thin, using the formula, 

\begin{equation} 
M_{dust}=\frac{F_{\nu}D^2}{\kappa _{\nu}B_{\nu}(T_{d})}, \label{dustmass}
\end{equation}

where we have adopted a value of the mass opacity, 
$\kappa _{\nu }=0.1 \left(\frac{250\mu m}{\lambda _{92 GHz}} \right) ^\beta$ cm$^{2}$ g$^{-1}$ \citep{Hildebrand83} and  
$\beta $=2.
\citet{Chini97} estimated the value of $\beta$ to be 2 by the SED fitting toward FIR 1 and FIR 2 in the OMC-2 region. Since FIR 4 locates on the same molecular filament of OMC-2 as FIR 1 and FIR 2, we consider that $\beta$ = 2 is probably acceptable for FIR 4.  For the dust temperature we adopted $T_d$=52.5 K, which is derived from the $^{12}$CO ($J$=3--2) emission (see $\S$ 3.1).  The mean gas density ($\equiv$ $n$) in these dusty cores was derived by assuming a spherically-symmetric shape as follows; 

\begin{equation}
n=\frac{M_{dust}}{\frac{4}{3}\pi\biggl(\sqrt{\frac{D_{maj}}{2}\times \frac{D_{min}}{2}}\biggl)^{3}}, \label{density}
\end{equation}

where $D_{maj}$ and $D_{min}$ are the deconvolved size along the major and minor axis. The typical size, mass and the density of these cores are estimated to be $\sim$ 5 $\arcsec$ (= 2250 AU), 0.78 $M_{\odot}$, and 1.6 $\times$ 10$^7$ cm$^{-3}$, respectively. The total mass of FIR 4 clump estimated from the 3.3-mm dust continuum emission is 9.4 M$_{\odot}$, which is consistent with that estimated from the H$^{13}$CO$^{+}$ emission (10.5 M$_{\odot}$). The average density of FIR 4 clump is 9.8 $\times$$10^5$ cm$^{-3}$, assuming the spherically symmetric core with a radius of 15$\arcsec$ (6750 AU). Table \ref{Ident} summarizes these physical properties of the identified cores.

From the naturally- and uniformly-weighted interferometric 3.3 mm flux in FIR 4, the 850 $\mu$m flux is estimated to be 1.2 and 0.58 Jy, respectively, on the assumption of $\beta$ = 2, while the 850 $\mu$m flux measured with JCMT is 7.5 Jy \citep{Joho99}. Then, $\sim$ 84 and 92 \% of the total flux are missed in the naturally- and uniformly-weighted interferometric 3.3 mm continuum images. Then, the estimated mass of these cores ($\equiv$ M$_{dust}$) should be considered as lower limits. In Table \ref{Ident}, we also list the mass of these cores corrected for the effect of the missing flux ($\equiv$ M$_{cor}$), which can be considered as upper limits of the mass.

%% information to the copy editor.  This information will appear as a
%% footnote on the printed copy for the manuscript style file.  Nothing will
%% appear on the printed copy if the preprint or
%% preprint2 style files are used.

%% The eqnarray environment produces multi-line display math. The end of
%% each line is marked with a \\. Lines will be numbered unless the \\
%% is preceded by a \nonumber command.
%% Alignment points are marked by ampersands (&). There should be two
%% ampersands (&) per line.

%% Putting eqnarrays or equations inside the mathletters environment groups
%% the enclosed equations by letter. For instance, the eqnarray below, instead
%% of being numbered, say, (4) and (5), would be numbered (4a) and (4b).
%% LaTeX the paper and look at the output to see the results.

%% This section contains more display math examples, including unnumbered
%% equations (displaymath environment). The last paragraph includes some
%% examples of in-line math featuring a couple of the AASTeX symbol macros.

\section{Discussion}
\subsection{Interaction between the molecular outflow driven by FIR 3 \\
and FIR 4 clump}

 Several pieces of evidence for the interaction between the molecular outflow driven by FIR 3 and FIR 4 
clump were found in our observations. The extent of the SW outflow lobe from FIR 3 observed in the CO 
(1--0, 3--2) lines ($\sim$ 0.09 pc) is shorter than that of the NE lobe ($\sim$ 0.19 pc). 
This suggests that the propagation of the SW outflow is hampered by the material at the head of the outflow, 
and in fact FIR 4 clump locates at the tip of the south-western outflow.  
These pieces of morphological evidence suggest that the south-western outflow component is dammed by FIR 4 clump. 
We also detected the SiO and submillimeter CH$_{3}$OH emission at the tip of the south-western outflow. 
These molecular lines are often observed toward shocked regions, 
caused by the interaction between outflows and dense gas \citep{Avery96,Bach01}. 
Hence, the detection of the SiO and submillimeter CH$_{3}$OH emission in FIR 4 is probably the chemical 
evidence for the interaction between the molecular outflow and the dense gas. 
In addition, the line width of the H$^{13}$CO$^{+}$ emission at FIR 4 (dV$_{FWHM}$ $\sim$ 1.1 km s$^{-1}$) is 
larger than the average value in other OMC-2/3 cores 
\citep[dV$_{FWHM}$ $\sim$ 0.8 km s$^{-1}$;][]{ike07}
which may suggest the larger turbulence due to the interaction with the outflow.

\begin{figure}
  \includegraphics[width=15cm,clip]{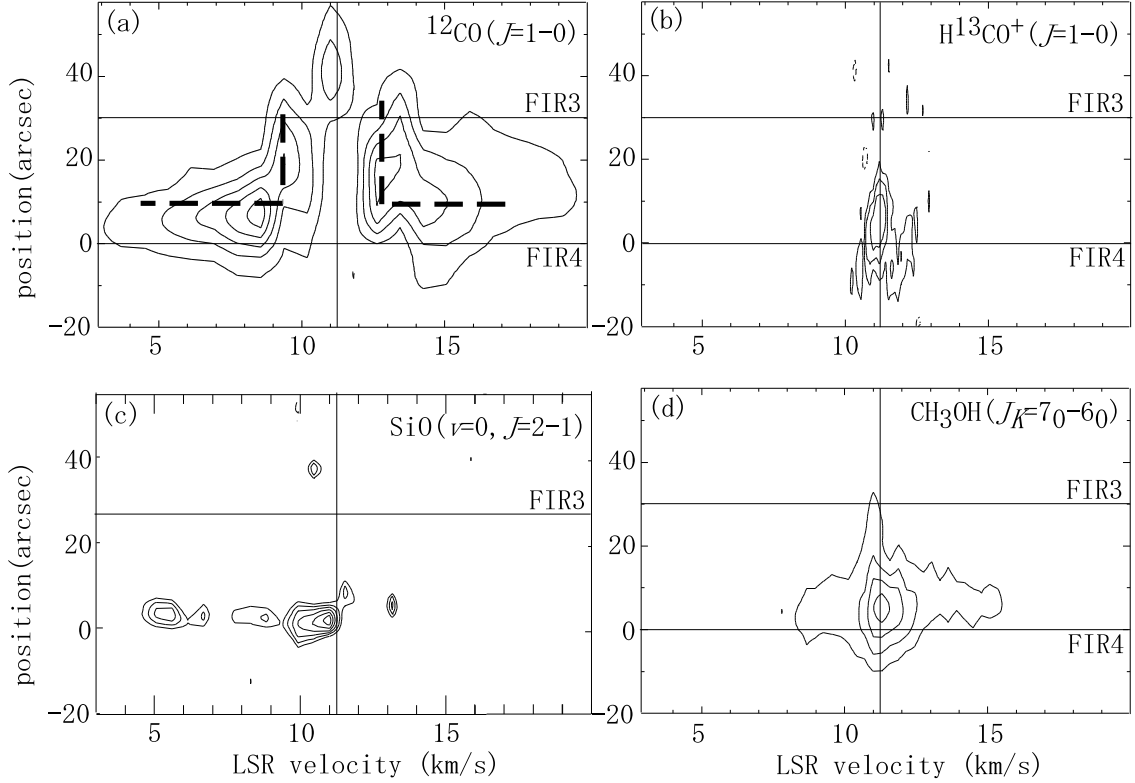}
  \caption{Position-Velocity (P-V) diagrams of the $^{12}$CO ($J$=1--0) (upper left), H$^{13}$CO$^{+}$ ($J$=1--0)
(upper right), SiO ($v$=0, $J$=2--1) (lower left) and CH$_3$OH ($J_K$=7$_0$--6$_0$) (lower right) emission in
the FIR 4 region.
Two horizontal lines in each panel show the position
of FIR 3 and FIR 4, while a vertical line the systemic velocity of 11.3 km s$^{-1}$ in FIR 4 estimated from 
the H$^{13}$CO$^{+}$ data. The cut line of
panel (a), (b), and (d) is adopted to be along P.A. = 30$^\circ$, passing through the position of FIR 3 and FIR 4.
The cut line of panel (c) is taken to be along P.A. = 35$^\circ$, passing through the compact SiO emission
and FIR 3.
Contour levels in panel (a) and (d) start at $\pm$ 5 $\sigma$ levels with an interval of 5 $\sigma$. 
Contour levels in panel (b) start at $\pm$ 3 $\sigma$ levels with an interval of 3 $\sigma$. 
Contour levels in panel (c) start at $\pm$ 2 $\sigma$ levels with an interval of 1 $\sigma$. 
The rms noise level (1 $\sigma$) in panel (a)-(d)
are 0.57 Jy beam$^{-1}$, 0.12 Jy beam$^{-1}$, 0.2 Jy beam$^{-1}$, and 0.17K, respectively. 
}
  \label{PV}
\end{figure}

 Figure \ref{PV}a-d show Position-Velocity (P-V) diagrams in the $^{12}$CO, H$^{13}$CO$^{+}$, SiO, and 
the CH$_{3}$OH emission. The P-V diagrams in the $^{12}$CO, H$^{13}$CO$^{+}$, SiO, and the CH$_{3}$OH 
emission show successive change of the line width along the direction from FIR 3 to FIR 4.  
First, the $^{12}$CO (1--0) emission shows a distinct L-shaped structure in the P-V diagram, 
and the location of the line broadening is at the most upstream. 
Second, the CH$_{3}$OH emission also shows a similar L-shape with a less broadening, 
which locates slightly at the downstream from the CO L-shape. 
The SiO emission shows the most extreme blueshifted components, 
which locates at the peak of the blueshifted CH$_3$OH component. Finally, the dense gas component traced by the H$^{13}$CO$^{+}$ emission locates at the most downstream of the successive emission distribution and is pinched by the L-shaped CO emission distribution in the P-V diagram. These successive emission distributions and the velocity structures are likely to trace the detail of the interaction. 
 
These observational results suggest an interconnection among the molecular outflow, dense gas, and the shock traced by the SiO and CH$_{3}$OH emission. We consider that these pieces of evidence support the presence of the interaction between the molecular outflow and dense gas.

\subsection{Outflow-triggered star formation}

 In the last section, we suggest that there is an interaction between the outflow driven by FIR 3 and dense gas of FIR 4 clump. On the other hand, in $\S$ 3.4 we have demonstrated that FIR 4 clump consists of eleven dusty cores. From these results, we presume that the interaction causes the fragmentation of FIR 4 clump into these cores and eventually the formation of the next generation of the cluster members. Hereafter we will examine this possibility from consideration of the Jeans instability, time scale, and the virial analyses.

\subsubsection{Jeans instability}

 First, we will examine whether the fragmentation into the dusty cores can be caused by the Jeans instability, by comparing the average separation of the cores to the Jeans length. The average 3-dimensional separation ($\equiv$ $\Delta l$) among these cores can be estimated by the following formula \citep{Pen95};

\begin{equation} 
\Delta l=2(\frac{1}{\gamma ^{\frac{1}{3}}}-1)r,
\end{equation}

\begin{equation} 
\gamma =\frac{n_{core}V_{core}}{V_{clump}},
\end{equation}

where $\gamma$ , $r$, $n_{core}$, $V_{clump}$, and $V_{core}$ are the volume filling factor, the average radius of the cores, the number of cores, and the average volume of FIR4 clump and a single core, respectively. V$_{core}$ was estimated from the average of the projected size of the dusty cores (Table \ref{core}), and V$_{clump}$ was estimated from the radius of 15$\arcsec$, on the assumption of the 3-dimensional spherical symmetry (see eq. 6). The estimated V$_{core}$, V$_{clump}$, and $r$ are $3.1 \times 10^{46}$ cm$^{3}$, $6.5 \times 10^{48}$ cm$^{3}$, and $1.1 \times 10^3$ AU, respectively. Then, the average separation $\Delta l$ is estimated to be $\sim$ $5 \times10^3$ AU. 

 In order to estimate the Jeans length ($\equiv$ $\lambda_J$), we use the following equation \citep{Naka07};

\begin{equation} 
\lambda_J^2=\frac{\pi C_{eff}^2}{G\rho_0},
\end{equation}

where $C_{eff}$, G, and $\rho_0$ are the effective sound speed, gravitational constant, and the average density of FIR 4 clump, respectively. $C_{eff}$ is the effective sound speed in FIR 4 clump ($\sim$ 0.62 km s$^{-1}$) derived from equation (4), and $\rho_0$ is estimated to be 9.8 $\times$ 10$^{5}$ cm$^{-3}$ in $\S$ 3.4. With these values, the Jeans length $\lambda _J$ is estimated to be $\sim$ $13 \times 10^3$ AU, which is on the same order of the average separation of the dusty cores. Hence it is possible that the fragmentation of FIR 4 clump into cores was caused by the Jeans instability. The Jeans instability in FIR 4 clump may be triggered by the interaction with the outflow from FIR 3, and in the next section we will discuss whether the fragmentation was caused after the interaction between the outflow driven by FIR 3 and FIR 4 clump.

\subsubsection{Time scale of the fragmentation of FIR 4 clump into cores}

 We can estimate the time scale of the fragmentation on the assumption that the fragmentation time scale 
($\equiv$ $\tau _{fragmentation}$) is the sound crossing time; 

\begin{equation}
\tau _{fragmentation}=\frac{\Delta l}{C_{eff}}.
\end{equation}

Then $\tau _{fragmentation}$ is estimated to be $\sim$ $3.8 \times 10^4 $ yr. 
Next, we estimate the time scale of the interaction between the outflow and dense gas, 
on the assumption that the interaction time scale $\tau _{interaction}$ is similar to 
the dynamical time $\tau_{d}$ of the north-eastern outflow driven by FIR 3, i.e. 
$\tau _{interaction}$ $\sim$ $\tau_{d}$. \citet{Takaha07} estimated $\tau_{d}$ to 
be $1.4 \times 10^4 $ yr.  
Then the interaction time scale $\tau _{interaction}$ is estimated to be $1.4 \times 10^4 $ yr. 
Therefore, the fragmentation time scale (3.8 $\times$ 10$^4$ yr) is similar to the interaction time scale 
(1.4 $\times$ 10$^4$ yr), and we suggest that the fragmentation was triggered by the 
interaction with the outflow from FIR 3. Moreover, the free-fall time in this region 
($\sim$ 1.3 $\times 10^4$ yr), which is estimated from the average density of each core 
($\sim$ 1.6 $\times$ 10$^7$ cm$^{-3}$), is also similar to the interaction time-scale, 
and hence it is possible that Class 0/I protostars are already formed in the course of the interaction \citet{nie03}. In fact, there exists an MIR source (MIR 24) identified as a Class 0 protostar, and it is possible that MIR 24 is formed in the course of the interaction.

We suggest that in this particular case of FIR 4 the interaction between the outflow and the dense gas most likely triggers the fragmentation into cores, while we cannot exclude other possibilities such as the induced fragmentation due to the interaction with the HII region (M43) \citep{Hester05} as well as spontaneous fragmentation due to the gravity \citep{Inu97} and turbulence \citep{Li04}.

\subsubsection{Outflow-triggered star formation}

Finally, in order to examine whether these dusty cores have a potential to form protostars inside, we compare the H$_2$ mass of the cores estimated from the 3.3 mm dust-continuum flux and the virial mass. The virial mass ($\equiv$ M$_{virial}$) is estimated by the following equation;

\begin{equation}
M_{virial}=\frac{5rC_{eff}^2}{2G},
\end{equation}

where $r$, $C_{eff}$ and G are the radius of the core, the effective sound speed (equation 4) and the gravitational constant, respectively. Here, we assume that the H$^{13}$CO$^{+}$ emission is associated with all the dusty cores. Then we adopted the H$^{13}$CO$^{+}$ line width ($\Delta V_{FWHM}$; see Table \ref{Ident}) at each peak of the dusty core as the line width in the dusty cores. The estimated virial masses are within the range of the expected H$_2$ mass of these cores, that is, from M$_{dust}$ to M$_{cor}$ (Table \ref{Ident}). These results suggest that the observed dusty cores have a potential to form stars.

\begin{figure}
  \includegraphics[width=10cm,clip]{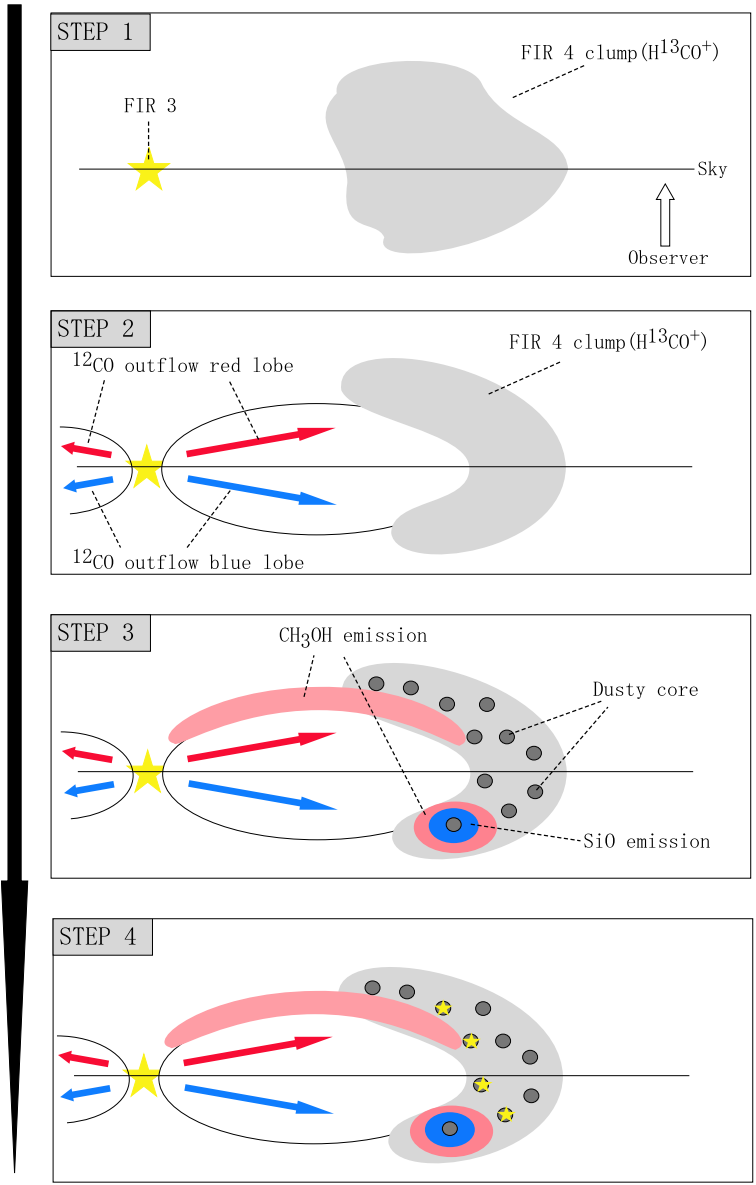}
  \caption{Schematic picture of the outflow-triggered star formation in the FIR 4 region.
In STEP 1, FIR 3 was born and drove the outflow. In STEP 2, the outflow driven by FIR 3 interacted
with FIR 4 clump. In STEP 3, the interaction between the outflow and FIR 4 clump caused the fragmentation
of FIR 4 clump into cores. In the final STEP, these cores form stars. The current stage of the FIR 4 region may be
between STEP 3 and STEP 4. In these figures, horizontal lines show the plane of the sky and the observers are
located at the bottom.}
  \label{scenario}
\end{figure}

From the above considerations, we suggest that the interaction between FIR 4 clump and the outflow driven by FIR 3 caused the fragmentation of FIR 4 clump into cores and triggered the next generation of cluster formation. Our scenario of the star formation in the FIR 4 region is summarized in Figure \ref{scenario} as;

[STEP 1]: FIR 3 was born and drove the outflow; 

[STEP 2]: The outflow driven by FIR 3 started interacting with FIR 4 clump; 

[STEP 3]: The interaction caused the fragmentation of FIR 4 clump into cores; 

[STEP 4]: These cores will form stars. 

The present stage of the star formation in the FIR 4 region may be between STEP 3 and STEP 4. 
We speculate that dusty cores will form protostars inside, and eventually, next generation of the stellar cluster in the FIR 4 region. 

Possible outflow-triggered star formation has also been reported observationally in NGC 1333 \citep{San01}, L1551 NE \citep{Yoko03}, and NGC 2264 IRS 1 \citep{Nakano03}. In NGC 1333, \citet{San01} found that one of the cores in the 850 $\mu$m emission (SK-1) locates at the tip of the shell evacuated by the outflow, and they suggest that SK-1 is a possible example of the outflow-triggered star formation. In L 1551 NE, \citet{Yoko03} have found an arc-shape structure of dense gas open toward South-West, or the direction of the molecular outflow driven by L1551 IRS 5, in the CS (3--2) emission. They interpret that the dense gas in L1551 NE is affected by the outflow from L1551 IRS 5, creating an arc-shape, and the protostellar formation of L1551 NE. In NGC 2264 IRS 1, \citet{Nakano03} have found that a dense-gas shell seen in the H$^{13}$CO$^{+}$ emission locates around the outflow driven by IRS 1, and at the inner edge of the dense shell there are three compact cores. They suggest that these cores are entrained or compressed material formed by the outflow and that these cores may represent future sites of the formation of a group of the low-mass stars. Our observational studies in the OMC-2 FIR 3/4 region have suggested that the interaction between the outflow and dense gas can cause the formation of stellar clusters. Recent 3-dimensional MHD simulations by \citet{Naka07} suggest that protostellar outflows can trigger cluster formation through shock compression, providing a theoretical support to our scenario.

\section{SUMMARY}
 We have carried out high angular-resolution ($\sim$ 3$\arcsec$ - 7$\arcsec$) millimeter interferometric observations of OMC-2 FIR 3/4 with the NMA in the H$^{13}$CO$^{+}$ ($J$=1--0), $^{12}$CO ($J$=1--0), CS ($J$=2--1), and SiO ($v$=0, $J$=2--1) lines as well as in the 3.3 mm continuum emission, and submillimeter single-dish observations with ASTE in the $^{12}$CO ($J$=3--2) and CH$_3$OH ($J_K$=$7_K$--$6_K$) lines. The main results of our new millimeter and submillimeter observations are summarized as follows:

\begin{enumerate}
\item We suggest that the outflow driven by FIR 3 interacts with the 0.07 pc-scale dense gas associated with FIR 4 ($\equiv$ FIR 4 clump), from morphological, kinematical, chemical and physical evidence. In the morphological evidence, we found that the length of the SW outflow lobe from FIR 3 observed in the CO (1--0, 3--2) lines ($\sim$ 0.09 pc) is shorter than that of the NE lobe ($\sim$ 0.19 pc), and FIR 4 clump is located at the tip of the SW molecular outflow. These results suggest that the SW lobe has been dammed by the interaction with FIR 4 clump. In the chemical and physical evidence, we detected the SiO ($v$=0, $J$=2-1) emission and the submillimeter CH$_3$OH ($J_K$=7$_K$-6$_K$; K=-1,0,2) emission around FIR 4 clump, which are known to trace shocks caused by the interaction between outflows and dense cores. In the kinematical evidence, the P-V diagrams in the $^{12}$CO, CH$_{3}$OH, SiO, and H$^{13}$CO$^{+}$ emission show a successive change of the line width along the direction from FIR 3 to FIR 4. The $^{12}$CO (1--0) emission show a distinct L-shaped structure in the P-V diagram, and the location of the broadening is at the most upstream. The CH$_{3}$OH emission also shows a similar L-shape at the downstream from the CO L-shape, and the SiO emission shows the most blueshifted component at the peak of the blueshifted CH$_3$OH component. The dense gas component traced by the H$^{13}$CO$^{+}$ emission locates at the most downstream of the successive emission distribution and is pinched by the L-shaped CO emission distribution in the P-V diagram. These results indicate the presence of the interaction between the outflow and dense gas in FIR 4.

\item In the 3.3-mm continuum emission, we have first resolved FIR 4 clump into eleven dusty cores with a size of 1500 - 4000 AU and a mass of 0.2 - 1.4 M$_{\odot}$. The 3-dementinal separation among these cores, 5000 AU, is on the same order of the Jeans length, 13000 AU. Moreover, the estimated time scale of the fragmentation into the cores, 3.8 $\times10^4$ yr, is also similar to the time scale of the interaction between FIR 4 clump and the outflow from FIR 3, that is, 1.4 $\times 10^4$ yr. Therefore, it is possible that the fragmentation of FIR 4 clump into the cores has been caused by the Jeans instability triggered by the interaction with the outflow driven by FIR 3. We suggest that in this particular case of FIR 4 the outflow interaction most likely triggers the fragmentation into cores.

\item We speculate that the dusty cores in FIR 4 will form protostellar sources eventually, and that the FIR 4 region is in the course of cluster formation. We suggest that the outflow driven by FIR 3 triggered the next generation of the cluster formation in the FIR 4 region, and that the FIR 4 region is one of the promising samples of the outflow-triggered cluster formation.

\end{enumerate}

%% The displaymath environment will produce the same sort of equation as
%% the equation environment, except that the equation will not be numbered
%% by LaTeX.


%% Included in this acknowledgments section are examples of the
%% AASTeX hypertext markup commands. Use \url without the optional [HREF]
%% argument when you want to print the url directly in the text. Otherwise,
%% use either \url or \anchor, with the HREF as the first argument and the
%% text to be printed in the second.

\acknowledgments
We are grateful to the staffs at the Nobeyama Radio Observatory (NRO) for both operating the
NMA and helping us with the data reduction, and to M. Yamada, Y. Kurono, and T. Tsukagoshi
for their helpful comments. NRO is a branch of the National Astronomical Observatory, National Institutes of Natural Sciences, Japan. We also acknowledge the ASTE staffs for both operating ASTE and helping us with the data reduction. We would like to thank the anonymous referee for providing helpful suggestions to improve the paper. Observations with ASTE were (in part) carried out remotely from Japan by using NTT's GEMnet2 and its partner R\&E (Research and Education) networks, which are based on AccessNova collaboration of University of Chile, NTT Laboratories, and National Astronomical Observatory of Japan. This work was supported by Grant-in-Aid for Scientific Research A 18204017.


%% in parentheses, after the keyword, but they will not be verified.



\appendix




\clearpage

\begin{deluxetable}{lcccc}
\tablecolumns{5}
\tabletypesize{\scriptsize}
\rotate
\tablecaption{Parameters for the NMA observations}
\tablewidth{0pt}
\tablehead{
\colhead{parameter} & \colhead{H$^{13}$CO$^{+}$ ($J$=1--0)} & \colhead{$^{12}$CO ($J$=1--0)} & \colhead{SiO ($v$=0, $J$=2--1)} & \colhead{CS ($J$=2--1)}
}
\startdata
Configuration$^a$ & D, C and AB & D & D & D, C and AB \\
 Baseline [k$\lambda$] & 2.9-115 & 3.0-30.8  & 2.9-23.4 & 3.2-113.6 \\
 Phase reference center (J2000) & \multicolumn{4}{c}{$\alpha _{J2000}=05^h 35^m 26$\fs$47, \delta _{J2000}=-05^{\circ} 10\arcmin 00\farcs4$} \\
 Primary beam HPBW [arcsec] &  77$\arcsec$ & 62$\arcsec$ & 77$\arcsec$  & 70$\arcsec$ \\
 Synthesized Beam HPBW [arcsec] & 9$\farcs$54$\times $5$\farcs$15 & 6$\farcs$59$\times$ 5$\farcs$79 & 10$\farcs$22 $\times$ 6$\farcs$37  & 6$\farcs$68 $\times$ 4$\farcs$00 \\
 Velocity resolution [km s$^{-1}$ ]& 0.108 km s$^{-1}$  & 0.406 km s$^{-1}$  & 0.539 km s$^{-1}$ & 48.94 km s$^{-1}$ \\
 Gain calibrator$^b$ & \multicolumn{4}{c}{0528+134}\\
 Bandpass calibrator$^c$ & 3C273, 0420-014, 3C84, 0528+134 & 3C273, 3C84 & 3C273, 3C84 & 3C273, 0420-014, 3C84, 0528+134 \\
 System temperature in DSB [K]$^d$ & 100-400 K & 150-300 K & 100-350 K & 100-400 K \\
 Rms noise level [Jy beam $^{-1}$] & 1.2$\times $10$^{-1}$ Jy beam $^{-1}$ & 5.7$\times $10$^{-1}$ Jy beam $^{-1}$ & 2.0$\times $10$^{-1}$ Jy beam $^{-1}$ & 5.1$\times $10$^{-3}$ Jy beam $^{-1}$ \\

\enddata
\label{NMAobs}
\tablenotetext{a}{D and AB are the most compact and sparse configurations, respectively.}
\tablenotetext{b}{A gain calibrator, 0528+134, was observed every 20 minutes.}
\tablenotetext{c}{The Bandpass calibrations were archived by 40 minute observations of each calibrator.}
\tablenotetext{d}{The system noise temperature of the SIS receiver was measured toward the zenith.}
\end{deluxetable}

\clearpage

\begin{deluxetable}{lcc}
\tabletypesize{\scriptsize}
\tablecaption{Parameters for the continuum observations}
\tablewidth{0pt}
\tablehead{
\colhead{parameter} & \colhead{Fig.\ref{continuum}a} & \colhead{Fig.\ref{continuum}b} 
}
\startdata
Baseline [k$\lambda$] & 2.9-115 & 2.9-115  \\
Weighting & Natural & Uniform \\
Beamsize (HPBW) [arcsec] & 6$\farcs$96$\times $ 4$\farcs$03 (P.A.=-34.59$^\circ$)& 6$\farcs$55$\times $3$\farcs$34 (P.A.=-40.31$^\circ$) \\
P.A. of the beam [$^\circ$] & -34.59 & -40.31\\
Rms noise level [Jy beam$^{-1}$] & 1.2$\times $10$^{-3}$ & 1.4$\times $10$^{-3}$  \\
\enddata
\label{continuumobs}
\end{deluxetable}

\clearpage
\begin{deluxetable}{lccccc}
\tabletypesize{\scriptsize}
\tablecaption{Parameters of FIR 4 clump derived from the H$^{13}$CO$^{+}$ ($J$=1--0) emission}
\tablewidth{0pt}
\tablehead{
\colhead{parameter} &
\colhead{M$_{gas}$ } & 
\colhead{V$_{sys}$ } & 
\colhead{$\Delta $V$_{FWHM}$}& 
\colhead{P$_{gas}$ } & 
\colhead{E$_{gas}$ } 
\\
\colhead{} &
\colhead{[M$_{\odot}$]} & 
\colhead{[km s$^{-1}$]} & 
\colhead{[km s$^{-1}]$}& 
\colhead{[M$_{\odot}$ km s$^{-1}$]} & 
\colhead{[M$_{\odot}$ km$^2$ s$^{-2}$]}
}
\startdata
value & 10.5 & 11.3 & 1.1 & 10.5 & 5.2  \\
\enddata
\label{Physical_qua}
\end{deluxetable}

\clearpage

\begin{deluxetable}{lcccccccccc}
\scriptsize
\tabletypesize{\scriptsize}
\rotate
\tablecaption{Identified dusty cores}
\tablewidth{0pt}
\tablehead{
\colhead{source} 
&\colhead{$\alpha_{J2000}$ } 
&\colhead{$\beta _{J2000}$} 
&\colhead{D$_{maj} \times$ D$_{min}$$$$^b$} 
&\colhead{P.A.} 
&\colhead{M$_{dust}$$^c$} 
&\colhead{M$_{cor}$$^d$ } 
&\colhead{$n$$$}
&\colhead{M$_{virial}$} 
&\colhead{$\Delta $V$_{FWHM}$} 
&\colhead{$C_{eff}$ }
\\
\colhead{} 
&\colhead{} 
&\colhead{} 
&\colhead{[$\arcsec \times \arcsec$]} 
&\colhead{[$\arcsec$]} 
&\colhead{[M$_{\odot}$]} 
&\colhead{[M$_{\odot}$] } 
&\colhead{[$\times$10$^{7}$ cm$^{-3}$]}
&\colhead{[M$_{\odot}$]} 
&\colhead{[km s$^{-1}$]} 
&\colhead{[km s$^{-1}$]}
}
\startdata
         FIR 3  & 05$^h$ 35$^m$ 27$\fs$6 & -05$^\circ$ 09$\arcmin$ 34$\farcs$0 & 11.7$\times$3.6 & 140.46 & 3.0 & 18.7 & 3.1 & \nodata & \nodata  & \nodata \\                                        
         FIR 4a$^{a}$ & 05 35 27.4 & -05 09 49.0 & 5.4$\times$3.5 & 34.05 & 0.51 & 3.2 & 1.8 & 1.1 & 0.53 & 0.47\\
         FIR 4b$^{a}$ & 05 35 26.6 & -05 09 51.0 & 6.3$\times$3.9 & 150.20 & 1.1 & 6.9 & 2.6 & 2.6 & 1.28 & 0.68\\
         FIR 4c & 05 35 27.0 & -05 09 54.0 & 9.0$\times$4.0 & 168.88 & 0.45 & 5.9 & 0.59 & 1.6 & 0.57 & 0.48\\
         FIR 4d & 05 35 26.2 & -05 09 55.9 & 4.7$\times$4.4 & 139.87 & 0.76 & 9.6 & 2.2 & 1.2 & 0.55 & 0.47\\
         FIR 4e & 05 35 26.9 & -05 09 57.5 & 8.4$\times$5.7 & 4.302  & 1.0 & 14.3 & 0.93 & 3.8 & 1.31 & 0.69\\
         FIR 4f & 05 35 26.6 & -05 09 58.9 & 6.1$\times$3.1 & 168.26 & 0.6 & 7.8 & 2.0 & 1.7 & 0.98 & 0.59\\
         FIR 4g$^{a}$  & 05 35 26.6 & -05 09 59.4 & 8.3$\times$5.0 & 179.53 & 1.2 & 7.5 & 1.3 & 2.1 & 0.81 & 0.54\\
         FIR 4h & 05 35 27.1 & -05 10 04.5 & 7.1$\times$5.7 & 156.55 & 0.64 & 8.3 & 0.70 & 3.3 & 1.25 & 0.67\\
         FIR 4i & 05 35 26.5 & -05 10 06.0 & 6.5$\times$4.8 & 162.23 & 0.21 & 2.7 & 0.34 & 1.7  & 0.70 & 0.50\\
         FIR 4j$^{a}$  & 05 35 27.2 & -05 10 08.0 & 5.6$\times$4.0 & 111.31 & 1.38 & 8.8 & 3.7 & 1.7 & 0.90 & 0.56 \\
         FIR 4k$^{a}$ & 05 35 26.3 & -05 10 16.4 & 6.35$\times$3.91 & 165.76 & 0.67 & 4.2 & 1.5 & 2.1 & 1.07 & 0.61\\  
         FIR 4 clump & \nodata  & \nodata  & \nodata  & \nodata   & 9.4 & 58.6 &  0.098$^{e}$ &  \nodata   & \nodata  &  \nodata\\ 
\enddata
\label{core}
\label{Ident}
\tablenotetext{a}{Identified in the naturally-weighted map.}
\tablenotetext{b}{The FWHM size estimated by the 2-dimensional Gaussian fitting to the image.}
\tablenotetext{c}{Derived from the total flux density integrated within the 3$\sigma$ contour levels,
with $D$=450 pc, $\beta $=2, and $T_{d}$=52.5 K. See texts for details.}
\tablenotetext{d}{H$_2$ mass corrected for the effect of the missing flux. See texts for details.}
\tablenotetext{e}{Assuming a spherically-symmetric core with a radius of 15$\arcsec$ (=6750 AU).}

\end{deluxetable}

\clearpage

\end{document}